\documentclass[a4paper]{article}

\usepackage[english]{babel}
\usepackage[ansinew]{inputenc}
\usepackage{amsmath}
\usepackage{amssymb}
\usepackage{stmaryrd}
\usepackage{mathrsfs} 
\usepackage{enumerate}
\usepackage{dsfont}
\usepackage{amsthm}
\usepackage{geometry}
\usepackage{accents}
\usepackage{amsxtra}
\usepackage{amsgen,amstext,amsbsy,amsopn}

\newcommand{\version}{March 1, 2012}
\numberwithin{equation}{section}
\newcommand{\bdm}{\begin{displaymath}}
\newcommand{\edm}{\end{displaymath}}
\newcommand{\bdn}{\begin{eqnarray}}
\newcommand{\edn}{\end{eqnarray}}
\newcommand{\bay}{\begin{array}{c}}
\newcommand{\eay}{\end{array}}
\newcommand{\ben}{\begin{enumerate}}
\newcommand{\een}{\end{enumerate}}
\newcommand{\beq}{\begin{equation}}
\newcommand{\eeq}{\end{equation}}

\newcounter{remark}[section]

\numberwithin{equation}{section}
\begin{document}

\markboth{\scriptsize{Massless stringfields - PY - \version}}{\scriptsize{Massless stringfields - PY - \version}}

\title{Massless, String Localized Quantum Fields for Any Helicity}

\author{M. Plaschke, J. Yngvason	\\ \normalsize\it Fakult\"at f\"ur Physik, Universit{\"a}t Wien,	\\ \normalsize\it Boltzmanngasse 5, 1090 Vienna, Austria.}

\date{\version}

\maketitle

\begin{abstract} For any massless, irreducible representation of the covering of the proper, orthochronous Poincar\'e group we construct  covariant, free  quantum fields that generate the representation space from the vacuum and are localized in semi-infinite strings in the sense of commutation or anti-commutation of the field operators at space-like separation of the strings. 
Besides the space-like string direction the field operators carry a spinor or tensor index for one of the finite dimensional representations $D^{(\frac j2,\frac k2)}$ of $SL(2,\mathbb C)$. For given $(j,k)$ the possible integer or half-integer values that the helicity $h$ can take are for string fields only restricted by the condition $|h|\leq \frac{j+k}2$, in contrast to the case of point localized fields where $h=\frac k2-\frac j2$ must hold. For infinite helicity no additional index on the string-field is needed in the Bose case while in the Fermi case the fields carry an additional spinor index. For finite helicity we consider in particular string-localized fields that are generalized potentials for point localized fields. The short distance behavior of their two-point functions                                                                                                                                                                                                                                                                                                                                                                                                                                                                                                                                                                                                                                                                                                                                                                                                                                                                                                                                                                                                                                                                                                                                                                                                                                                                                                                                                                                                                                                                                                                                                                                                                                                                                                                                                                                                                                                                                                                                                                                                             is independent of the helicity.
	\vspace{0,2cm}

	\vspace{0,2cm}
	
	\end{abstract}

\section{Introduction}

The principle of locality is one of the most important in modern quantum field theory.
It states that observables are measurable in bounded space-time regions, and that measurements
of space-like separated observables are compatible. Typically one uses point-localized
quantum fields (more precisely, smeared field operators  localized in bounded
space-time regions). To implement locality it is required that field operators commute for space-like separated
arguments. But in the formalism of local quantum field theory \cite{haag} only the observables
have to be localized in bounded regions, the (in general unobservable) fields can
have different localization properties.

One possibility to generalize the notion of localization is to use string-localized
fields as studied in \cite{msy04,msy05}. The basis for their analysis is provided by the paper of Brunetti,
Guido and Longo \cite{bgl}, where  a connection between irreducible positive energy
representations of the Poincar\'e group and localization in space-like cones is established using the concept of modular localization. The localization regions of string-localized fields can be regarded as the cores of such space-like cones. A digression of the history of string-like objects in QFT, that can be traced back to a 1935 paper of P. Jordan \cite{jordan},  and an account of modular localization is given in the first two sections of \cite{msy05}.

The string-localized fields considered in \cite{msy04, msy05}  are free fields that generate upon application to the vacuum one of the following irreducible representations of the Poincar\'e group: Massive representations of integer spin, massless representations of infinite integer helicity, and the massless representation of helicity 1. The case of zero mass and helicity 2 has been treated in \cite{mund09} and massive representations of half-integer spin in \cite{oliveira, mundoliveira, reiter}. In the present paper we complete the picture by constructing free massless fields for any helicity. 

Covariant point-localized free fields for any mass, spin or helicity, except for the case of mass zero and infinite helicity, were constructed and analyzed already in the 1960's \cite{weinberg64a, weinberg64b,schwartz}.  These fields transform under the representations $D^{(\frac j2,\frac k2)}$ of $SL(2,\mathbb C)$ (see next section) and for group theoretical reasons the possible $(j,k)$ are restricted as follows: For massive representations of spin $s$ the condition $|\frac j2-\frac k2|\leq s\leq \frac j2+\frac k2$ must be fulfilled while in the massless case the condition for finite helicity $h$ is $h=\frac k2-\frac j2$. The massless representations of infinite helicity can not be generated by covariant point fields \cite{yngvason}. For string-fields,  on the other hand, no spinor index on the field operators is required in the massive case with integer spin and also not in the massless case with infinite helicity. In the massive case and half integer spin a spinor index for the 2-dimensional representation of $SL(2,\mathbb C)$ suffices \cite{oliveira, mundoliveira, reiter}. In the sequel we shall consider the restrictions for the massless case and any helicity and find for finite helicity $h$  the condition for a covariant transformation behavior w.r.t. a representation $D^{(\frac j2,\frac k2)}$ to be $|h|\leq \frac j2+\frac k2$. 

In particular, we shall define string-localized potentials for point localized fields of helicity $h$. For integer helicity $h$ the potentials transform with the representation $D^{(\frac{|h|}2,\frac{|h|}2)}$ while the point localized fields transform with $D^{(|h|,0)}$ or $D^{(0,|h|)}$. In the ultraviolet the two-point function of the field behaves like momentum to the power $2|h|$ while the behavior of the potentials is bounded independently of $h$. For half-integer $h$ the potentials can be chosen to transform w.r.t. $D^{(\frac{|h|}2+\frac 14,\frac{|h|}2-\frac 14)}$ resp. $D^{(\frac{|h|}2-\frac 14,\frac{|h|}2+\frac 14)}$.

For the representations with unbounded integer helicity string localized Bose fields were constructed in \cite {msy05,msy04}. We extend this construction to cover also the case of unbounded half-integer helicity. Here Fermi commutation relations and an additional spinor index on the field operator are required.

As a final remark we note that theories of (interacting, gauged) massless fields of helicity higher than two, initiated in particular by 
C. Fronsdal \cite{f} and Fradkin and Vassiliev \cite{fv}, have in the past years received considerable attention, see, e.g.\ the review \cite{bbs}. Whether the concept of string localized fields may turn out to be useful in this context is an open question.

\section{String-Localized Quantum Fields}
\subsection{Definitions}
For the finite dimensional representations of $SL(2,\mathds{C})$ (the universal covering group of the proper orthochronous Lorentz group $\mathcal{L}_+^\uparrow$) we use the convention and notation of \cite{wightman}. For $A \in SL(2,\mathds{C})$ the corresponding Lorentz transformation $\Lambda(A)$ is defined by the adjoint action of $SL(2,\mathds{C})$ on the Lie algebra $sl(2,\mathbb C)$ that is isomorphic to four dimensional Minkowski space via the correspondence 
\beq x=(x^0,x^1,x^2,x^3)\leftrightarrow \undertilde{x} = 
\begin{pmatrix} x^0+x^3 & x^1-ix^2 \\ x^1+ix^2 & x^0-x^3 \end{pmatrix}.\eeq
If $j, k \in \mathds{N}\cup\{0\}$ the representation $D^{(\frac{j}{2},\frac k{2})}(A)$ is defined as $A^{\otimes_s j}\otimes\bar{A}^{\otimes_s k}$ on $(\mathds{C}^2)^{\otimes_s j}\otimes(\mathds{C}^2)^{\otimes_s k}$, where $\otimes_s$ denotes the symmetrized tensor product and the bar complex conjugation. Then a spinor $\xi_{\alpha_1,...,\alpha_j,\dot{\beta}_1,...,\dot{\beta}_k}$ transforms according to
\begin{equation*}
\xi_{\alpha_1,...,\alpha_j,\dot{\beta}_1,...,\dot{\beta}_k} \mapsto \sum_{(\rho)(\dot{\sigma})} A_{\alpha_1\rho_1}...A_{\alpha_j\rho_j} \bar{A}_{\dot{\beta}_1\dot{\sigma}_1}...\bar{A}_{\dot{\beta}_k\dot{\sigma}_k}\ \xi_{\rho_1,...,\rho_j,\dot{\sigma}_1,...\dot{\sigma}_k}\, .
\end{equation*}
We choose the Minkowski metric to have  signature $(1,-1,-1,-1)$. Denoting by $H^3:= \{e\in\mathbb{R}^4: e\cdot e=-1\}$ the hyperboloid of space-like directions and by $\alpha$ a collection of dotted or undotted spinor indices, a string-localized quantum field is an operator valued tempered distribution $\phi_\alpha(x,e)$ over $\mathds{R}^4\times H^3$, acting on a Hilbert space $\mathcal{H}$ and satisfying the following assumptions:
\begin{itemize}
\item \emph{String-(anti-)Locality}: If the strings $x_1 + \mathds{R}^+ e'_1$ and $x_2 + \mathds{R}^+ e_2$ are space-like separated for all $e'_1$ in an open neighborhood of $e_1$, then the fields $\phi_\alpha(x_1,e_1), \phi_\beta^\sharp(x_2,e_2)$ (anti-)commute, i.e.
\begin{equation}
[\phi_\alpha(x_1,e_1),\phi_\beta^\sharp(x_2,e_2)]_\mp = 0, \label{fieldlocality}
\end{equation}
where $\phi^\sharp$ either stands for the field $\phi$ itself or its adjoint $\phi^*$.
\item \emph{Covariance}: The field transforms covariantly under a unitary representation $U$ of the covering $\tilde{\mathcal{P}}_+^\uparrow$ of the proper, orthochronous Poincar\'e group, i.e.
\begin{equation}
U(a,A) \phi_\alpha(x,e) U(a,A)^{-1} = \sum_{\alpha '} D^{(\frac j2,\frac k2)}(A^{-1})_{\alpha\alpha '}\, \phi_{\alpha '}(\Lambda(A)x + a,\Lambda(A)e) . \label{covariance}
\end{equation}
The adjoint $\phi^*$ then  transforms according to the conjugate representation $D^{(\frac k{2},\frac{j}{2})}$.
\item \emph{Spectrum condition}: The joint spectrum of the generators of translations is contained in the forward lightcone $\overline{V}_+$ and there is a normalized, invariant vector $\Omega$ (vacuum), unique up to phase.
\end{itemize}
A further requirement that we shall make is the following: After smearing in $x$ with a testfunction $f$ the vector valued distribution
$e \mapsto \phi_\alpha^\sharp(f,e)\Omega$,
where $\Omega$ denotes the vacuum vector,  has an analytic continuation for $e$ in the forward tube $\mathcal{T}_+ := \mathds{R}^4+i V_+$ intersected with the complex hyperboloid $H^{3,c}$. This assumption is motivated by the free fields of \cite{msy05} and the analyticity properties of their intertwiners. Together with the spectrum condition it implies that the two-point function
\begin{equation*}
{\mathcal W}_{\alpha\beta}^{0,\sharp}(x, x'; e,e') := \langle\Omega, \phi_\alpha(x,e)\phi_\beta^\sharp(x',e')\Omega\rangle
\end{equation*}
has an analytic continuation to the domain
\begin{equation}\label{domain}
\begin{split}
&x' \in \mathcal{T}_+ , \ \ e' \in \mathcal{T}_+ \cap H^{3,c} \\
&x \in \mathcal{T}_- , \ \ e \in \mathcal{T}_- \cap H^{3,c},
\end{split}
\end{equation} with $ \mathcal{T}_{-}=\mathds{R}^4-i V_+$ the backward tube. An analogous statement holds  for 
\begin{equation*}
{\mathcal W}_{\beta\alpha}^{\sharp,0}(x' x; e',e) := \langle\Omega, \phi_\beta^\sharp(x',e')\phi_\alpha(x,e)\Omega\rangle.
\end{equation*}

\subsection{PCT and Locality}
We now show that the PCT condition
\begin{equation}
{\mathcal W}_{\alpha\beta}^{0,\sharp}(x,x',e,e') = \pm (-1)^J {\mathcal W}_{\beta\alpha}^{\sharp,0}(-x',-x,-e',-e), \label{pct}
\end{equation}
where the $\pm$ sign refers to bosonic or fermionic fields respectively and $J$ denotes the total number of undotted indices, is equivalent to (anti-)locality for the two-point function,
\begin{equation}
{\mathcal W}_{\alpha\beta}^{0,\sharp}(x,x',e,e') = \pm {\mathcal W}_{\beta\alpha}^{\sharp,0}(x',x,e',e), \ \  x+\mathds{R}_+ e \text{ and } x'+\mathds{R}_+ e' \text{ space-like separated}.
\end{equation}
The following proof uses similar arguments as in \cite{msy04}. 
First note that ${\mathcal W}_{\alpha\beta}^{0,\sharp}(x,x',e,e')$ has an analytic continuation to the domain \eqref{domain}
and $ {\mathcal W}_{\beta\alpha}^{\sharp,0}(x',x,e',e)$ has an analytic continuation into the domain with 
$\mathcal{T}_+$ and $\mathcal{T}_-$ interchanged.
Now consider $x,x',e$ and $e'$ such that the strings $x+\mathds{R}_+ e$ and $x'+\mathds{R}_+ e'$ are space-like separated. Then  there is a wedge $W$ such that $x +\mathbb{R}^+e \in W$ and $x'+\mathbb{R}^+e' \in W'$, where $W'$ denotes the causal complement of $W$ \cite[Appendix A]{msy05}. By translational invariance of the two-point function it can be assumed that the edge of $W$ contains the origin, such that $x,e\in W$, $x',e' \in W'$.
By covariance of the two-point function we get
\begin{equation}
\begin{split}
&{\mathcal W}^{0,\sharp}_{\alpha\beta}(x,-\Lambda(t)R(\pi)x',e,-\Lambda(t)R(\pi)e') \\ 
= \sum_{\alpha ',\beta '} D^{(0)}_{\alpha\alpha '}(A_{\Lambda}(t)A_R(\pi))&D^{(\sharp)}_{\beta\beta '}(A_{\Lambda}(t)A_R(\pi)) {\mathcal W}^{0,\sharp}_{\alpha '\beta '}(R(\pi)\Lambda(-t)x,-x',R(\pi)\Lambda(-t)e,-e'),
\end{split} \label{cov}
\end{equation}
where $D^{(0)}$ and $D^{(\sharp)}$ are the representations according to which the fields $\phi_\alpha$ and $\phi^\sharp_\beta$ transform. The matrices $A_{\Lambda}(t), A_R(\theta) \in SL(2,\mathds{C})$ denote boosts in the direction of the wedge $W$ and rotations parallel to the edge of the wedge respectively and are defined for the standard wedge $W_0=\{x:\ x^3\geq |x^0|\}$ according to
\begin{equation}\label{boostandrot}
A_{\Lambda}(t) = \begin{pmatrix} \exp ({t/2}) & 0 \\ 0 & \exp({-t/2}) \end{pmatrix}, \ \ \
A_{R}(\theta) = \begin{pmatrix} \exp({i\theta/2}) & 0 \\ 0 & \exp({-i\theta/2}) \end{pmatrix}.
\end{equation}
The transformations for a general wedge $W=\Lambda W_0$ are defined accordingly by adjoining the matrices for $W_0$ with an $A\in SL(2,\mathbb C)$ representing $\Lambda$.
$\Lambda(t)$ and $R(\theta)$ are the Lorentz transformations corresponding to $A_{\Lambda}(t)$ and  $A_R(\theta)$ respectively.

Now $A_{\Lambda}(t)$ is entire analytic in the boost parameter $t$  and for a complex $\tau$ in the strip $\mathbb{R} + i(0,\pi)$ the imaginary parts of $\Lambda(-\tau)x$ and $\Lambda(-\tau)e$  lie in $V_-$ while those of $\Lambda(\tau)e'$ and $ \Lambda(\tau)x'$ lie in $V_+$. Thus, by the analyticity of the two-point function, we can analytically continue the relation \eqref{cov} to $t \mapsto t+i\pi$ to get
\begin{equation}\label{relation}
{\mathcal W}_{\alpha\beta}^{0,\sharp}(x,x',e,e') = (-1)^J {\mathcal W}_{\alpha\beta}^{0,\sharp}(-x,-x',-e,-e'),
\end{equation}
where $J$ again denotes the total number of undotted indices. 
To justify the $(-1)^J$ sign consider again the matrices
\begin{equation*}
A_{\Lambda}(\pm i\pi) = \pm \begin{pmatrix} i & 0 \\ 0 & -i \end{pmatrix} = \pm A_R(\pi).
\end{equation*}
The representation $D^{(\frac{j}{2},\frac{k}{2})}(A_{\Lambda}(\pm i\pi)A_R(\pi))D^{(\frac{j'}{2},\frac{k'}{2})}(A_{\Lambda}(\pm i\pi)A_R(\pi))$  then yields exactly a factor of $(-1)^{j+j'}$ (where the matrix ``$D(A(i\pi))$'' must be evaluated by first considering $D(A(t))$ and then continuing to $t=i\pi$). Note that if the field transforms with  $D^{(\frac j2,\frac k2)}$ the adjoint field transforms with $D^{(\frac k2,\frac j2)}$, so $J$ is equal to $2j$ for the two point function of the field with itself, and $j+k$ for the \lq\lq mixed" functions containing $\phi$ and $\phi^*$. 

Assuming that the fields satisfy the PCT condition \eqref{pct} Eq.\ \eqref{relation} leads to
\begin{equation}
{\mathcal W}_{\alpha\beta}^{0,\sharp}(x,x',e,e') = \pm {\mathcal W}_{\beta\alpha}^{\sharp,0}(x',x,e',e),\label{locality}
\end{equation}
which is exactly the (anti-)locality. Conversely, assuming that the fields satisfy (anti-)locality leads to the PCT relation \eqref{pct} for $x +\mathbb{R}^+e \in W$ and $x'+\mathbb{R}^+e' \in W'$. Since both sides of \eqref{pct} are analytic in the domain \eqref{domain} the relation must hold everywhere.

Note also that as a consequence of \eqref{pct} and \eqref{locality} the two point functions $\langle\Omega, \phi_\alpha(x,e)\phi_\beta(x',e')\Omega\rangle$ and $\langle\Omega, \phi^*_\alpha(x,e)\phi_\beta^*(x',e')\Omega\rangle$ vanish for Fermi fields transforming with an irreducible representation $D^{(\frac j2,\frac k2)}$ and only the mixed functions containing $\phi$ and $\phi^*$ remain. The factor $(-1)^J$ is equal to -1 for the latter functions, because  $J=j+k$ is odd for Fermi fields.

A final point to note is that the PCT condition  \eqref{pct} implies the standard connection between spin and statistics: assuming Bose commutation relations for $j+k$ odd, or Fermi relations for $j+k$ even, leads in both cases to
\beq \langle\Omega, \phi_\alpha(x,e)\phi_\beta^{*}(x',e')\Omega\rangle+\langle\Omega, \phi_\beta^{*}(-x',-e')\phi_\alpha(-x,-e)\Omega\rangle=0
\eeq and thus $\phi_\alpha(x,e)\Omega=\phi_\alpha^{*}(x,e)\Omega\equiv 0$.

\subsection{Free Massless Fields for Finite Helicity}
From now on we will consider only fields with mass zero and the next sections will deal with the construction of free, massless string-localized fields for the representations of the Poincar\'e group of finite helicity $h$. Due to a version of the Jost-Schroer theorem for string-fields \cite{alazzawi,mund,steinmann} the fields are completely determined by the two point functions involving the field operators and their adjoints that, upon application to the vacuum, generate the one-particle space.
Because of the PCT-symmetry \eqref{pct} this space must contain states of both signs of the helicity. We thus make the the following ansatz for the field:
\begin{equation}
\phi_\alpha(x,e,h) = \int d\mu(p)\left[e^{ipx}u_\alpha(p,e,h)a^*(p,h) + e^{-ipx}v_\alpha(p,e,-h)a(p,-h)\right], \label{fieldansatz}
\end{equation}
with $d\mu({p})=\Theta(p^0)\delta(p\cdot p)d^4p$ the Lorentz invariant measure for mass zero,  $a^*(p,h)$, $a(p,h)$ the usual creation and annihilation operators, and yet undefined functions $u(p,e,h)$, $v(p,e,h)$.\\
The behavior of one-particle states under $SL(2,\mathbb C)$,
\begin{equation}
U(A) \Psi(p,h) = D^{(h)}(R(p,A))\Psi(\Lambda(A)^{-1}p,h)
\end{equation}
 with the `Wigner rotation' $R(p,A)$ (see \cite{msy05}, Eq.\ (19)) leads to the corresponding transformation properties of the creation and annihilation operators. In the massless, finite helicity  case $D^{(h)}(R(p,A))$ is just $e^{ih\theta(p,A)}$, where the angle $\theta$ is determined by the Wigner rotation. 
 
 For the field $\phi(x,e)$ to have the correct covariant transformation behavior \eqref{covariance}, the functions $u$ and $v$ have to satisfy the following intertwiner relations:
\begin{equation}\label{intertwinerrelation}
\begin{split}
D^{(\frac{j}{2},\frac k{2})}(A^{-1}) u(p,\Lambda(A) e,h) &= e^{ih\theta(p,A)} u(\Lambda(A)^{-1}p,e,h) \\
D^{(\frac{j}{2},\frac k{2})}(A^{-1}) v(p,\Lambda(A) e,h) &= e^{-ih\theta(p,A)} v(\Lambda(A)^{-1}p,e,h)
\end{split}
\end{equation}
Thus $u$ and $v$ intertwine the representation $D^{(\frac{j}{2},\frac k{2})}$ with the massless Wigner representations for helicity $+h$ and $-h$ respectively. Moreover, $u$ should have an analytic continuation into  $ \mathcal{T}_{+}\cap H^{3,c}$ and $v$ into $ \mathcal{T}_{-}\cap H^{3,c}$ . Then the PCT condition \eqref{pct} and hence (anti-)locality is equivalent to the condition
\begin{equation}\label{vfromu}
v(p,e,-h)=u(p,-e,h).
\end{equation}
In Section 3 we shall construct intertwiners satisfying Eq.\ \eqref{intertwinerrelation} and with the required analyticity properties in the string variable. It is enough to construct $u$ because then $v$ {\it defined} by \eqref{vfromu} has automatically the right properties if $u$ does.

\subsection{Self-adjoint Fields}
A free field defined as in \eqref{fieldansatz} is of course not self-adjoint in general. While the field $\phi_\alpha(x,e,h)$ creates one-particle states with helicity $h$, the adjoint field $\phi^*_{\dot{\beta}}(x,e,h)$ creates one-particle states with helicity $-h$ and transforms according to the conjugate representation (i.e. $D^{(\frac{k}{2},\frac{j}{2})}$ if the field transforms according to $D^{(\frac{j}{2},\frac k{2})}$). 

To obtain self-adjoint fields one way is to form the direct sum $D^{(\frac{j}{2},\frac k{2})}\oplus D^{(\frac{k}{2},\frac{j}{2})}$ and note that the real and imaginary parts of $\phi_\alpha$ together transform according to a real representation that is equivalent to this direct sum. Explicitly, acting with a unitary transformation on the direct sum
\begin{equation*}
\binom{\phi}{\phi^*},
\end{equation*}
one obtains
\begin{equation*}
\binom{\frac{1}{\sqrt 2}(\phi+\phi^*)}{\frac{1}{\sqrt 2i}(\phi-\phi^*)}
\end{equation*}
which transforms according to the real valued representation
\begin{equation*}
\begin{pmatrix} \text{Re}(D) & -\text{Im}(D) \\ \text{Im}(D) & \text{Re}(D) \end{pmatrix}.
\end{equation*}
This works for point fields and string fields alike. A well known example is the free photon field, where the field strength tensor $F_{\mu\nu}$ transforms according to the $D^{(1,0)}\oplus D^{(0,1)}$ representation. For $j=1, k=0$ the procedure leads to Majorana fields for helicity $\pm 1/2$.
 
For string fields there is, however, also another possiblity. If the field is bosonic we will see that for a given integer helicity $h$ one can always construct a field that transforms according to the representation $D^{(\frac{|h|}{2},\frac{|h|}{2})}$. In this case there are ``minimal intertwiners'', which satisfy $\overline{u(p,e,h)} = u(p,-e,-h)$ and are unique up to multiplication with a function $f(p\cdot e)$. 
Therefore, if we construct a field with these intertwiners and sum over positive and negative helicity, we can always obtain a self-adjoint field,
\begin{multline}
\phi_\alpha(x,e) \equiv \phi_\alpha(x,e,h) + \phi_\alpha(x,e,-h)\\ = \sum_{h=\pm|h|}\int d\mu(p)[e^{ipx}u_\alpha(p,e,h)a^*(p,h) + e^{-ipx} \overline{u_\alpha(p,e,h)}a(p,h)].
\end{multline}
Acting on the vacuum this field generates states of both helicities, $h$ and $-h$. 
In the fermionic case, i.e. for half-integer helicities, this construction is not possible because there is no ``symmetric'' representation $D^{(\frac{|h|}{2},\frac{|h|}{2})}$ for half-integer $h$. 

\section{Intertwiners for Massless Quantum Fields}
\subsection{Construction}
To solve the relations \eqref{intertwinerrelation} one first considers intertwiners for the standard momentum 
\begin{equation}
\hat p=(\hbox{$\frac12$} ,0,0,\hbox{$\frac12$})\quad \text{ i.e.}\quad \hat {\undertilde p}=
\begin{pmatrix} 1 & 0 \\ 0 & 0 \end{pmatrix}.
\end{equation}
The intertwiners for arbitrary momentum $p$ are fixed by those for $\hat p$ by the relation
\begin{equation}\label{upe}
u(p,e) = D^{(\frac{j}{2},\frac k{2})}(B_p)u(\hat p,\Lambda(B_p)^{-1}e),
\end{equation}
with the boost transformations $B_p\in SL(2,\mathbb C)$ satisfying $\Lambda(B_p) \hat p = p$. A possible choice is
\beq\label{boost}
B_p=\begin{pmatrix} \sqrt{p^0+p^3} & 0 \\ \frac {p^1+ip^2}{\sqrt{p^0+p^3}} & \frac 1{\sqrt{p^0+p^3}} \end{pmatrix}.\eeq
The little group, i.e., the stabilizer group of $\hat p$, is the double cover $\widetilde{E(2)}$ of the two-dimensional euclidean group and consists of the matrices
\beq\label{E2}
A_{z,\theta}=\begin{pmatrix} 1 & z \\ 0 & 1 \end{pmatrix}\begin{pmatrix} \exp({i\theta/2}) & 0 \\ 0 & \exp({-i\theta/2}) \end{pmatrix},
 \ z\in \mathds{C}, \ \theta\in [0,4\pi[.
\eeq

The intertwiners $u(\hat p,e) \equiv u(\undertilde{e})$ have to satisfy
\begin{equation}
D^{(\frac{j}{2},\frac k{2})}(A_{z,\theta}^{-1})\,u(A_{z,\theta}\undertilde{e}A_{z,\theta}^\dagger) = e^{ih\theta} u(\undertilde{e}), \label{intrel}
\end{equation}
where 
\begin{equation*}
\undertilde{e} = \begin{pmatrix} e^0+e^3 & e^1-ie^2 \\ e^1+ie^2 & e^0-e^3 \end{pmatrix}.
\end{equation*}
Such intertwiners can be constructed by setting
\begin{equation}\label{intertwinere}
u(\undertilde{e}) = \left[u_-^{\otimes m} \otimes_s u_+(\undertilde{e})^{\otimes j-m}\right]\otimes\left[{u_-}^{\otimes\bar{m}} \otimes_s \overline{u_+(\undertilde{e})}^{\otimes k-\bar{m}}\right], 
\end{equation}
where
\begin{equation}\label{upm}
u_- = \binom 1 0 , \hspace{1cm}
u_+(\undertilde{e}) = \undertilde{e}\binom 0 1 \end{equation}
and $\otimes_s$ denotes the symmetric tensor product. The intertwiners $u_-$ and $u_+(\undertilde{e})$ are the fundamental intertwiners for the representation $D^{(\frac{1}{2},0)}$. 

Noting that
\begin{equation}
\begin{split}
&A_{z,\theta}^{-1}\, u_- = \exp({-{i}\theta/2 })\, u_- , \\
&A_{z,\theta}^{-1}\, u_+(A_{z,\theta}\,\undertilde{e}\, A_{z,\theta}^\dagger) = 
\exp ({{i}\theta/2})\, u_+(\undertilde{e}) , \\
\end{split}
\end{equation}
the intertwiner \eqref{intertwinere} solves equation \eqref{intrel} if and only if the integers $0\leq m\leq j$ and $0\leq \bar{m}\leq k$ fulfill
\begin{equation}\label{hjk}
h = \left(\frac{j}{2}-m\right)-\left(\frac k{2}-\bar{m}\right).
\end{equation}
From these conditions follows that 
\beq\label{hrestriction} |h| \leq \frac{j+k}{2}\eeq
and this, besides the condition that $h$ is half integer if $j+k$ is odd and integer if $j+k$ is even,  is the only restriction on $h$ for given $(j,k)$. For point localized fields, on the other hand, the condition
is $h = \frac k2-\frac j2$ \cite{schwartz, weinberg}. This corresponds to the special case $m=j$ and $\bar m=k$, where the intertwiners \eqref{intertwinere} are independent of $\undertilde e$. 

The intertwiners given by \eqref{intertwinere} are linearly independent, because the fundamental intertwiners $u_-$ and $u_+(\undertilde{e})$ are. There are $(j+1)(k+1)$ of them for every fixed $j$ and $k$. They thus span the $(j+1)(k+1)$ dimensional representation space of $D^{(\frac{j}{2},\frac k{2})}$.  If $h$ satisfying \eqref{hrestriction} is given besides $(j,k)$ then the dimension $d$ of the corresponding subspace is the number of pairs $(m,\bar m)$ of integers fulfilling \eqref{hjk} as well as $0\leq m\leq j$ and $0\leq \bar{m}\leq k$. This number is
\begin{equation}\label{d}
d = \begin{cases} 
\frac{j+k}{2}-|h|+1 , & \text{if }\frac{j+k}{2}-|h|\leq \min\{j,k\} \\
\min\{j,k\}+1 , & \text{otherwise}
\end{cases}
\end{equation}
The coefficients of an expansion into the basis elements \eqref{intertwinere} can be written as functions $f_i(\hat p\cdot e)$ that should be analytic in the upper half plane and with at most polynomial growth at infinity and an inverse power of the imaginary part when approaching the real axis to comply with the analyticity and distributional requirements of the intertwiners. For a general $p$ the corresponding coefficients are then $f_i(p\cdot e)$. 

Summing up the preceding discussion, we have found intertwiners of the form
\beq\label{gerealintertwiner}
u(p,e)=\sum_{i=1}^{d}f_i(p\cdot e)u_i(p,e)
\eeq
where the $u_i(p,e)$ are given by \eqref{upe} with $u_i(\hat p,e)\equiv u_i(\undertilde e)$ taken from the basis \eqref{intertwinere}, $(m,\bar m)$ fulfilling  \eqref{hjk}, and $f_i$ having the stated analyticity and growth properties.\\

Like in \cite{msy05}, Proposition 3.4, the case where the intertwiners are analytic in $e$ in the whole complexified hyperboloid corresponds to point localized fields, i.e., they (anti-)commute for space like separated space-time points independently of the string directions. Genuinly string localized fields can be obtained by choosing functions $f_i$ with singularities on the real axis or in the lower half plane.

\subsection{Uniqueness}
We now want to show that the intertwiners $u(\undertilde e)$ constructed above are unique, up to multiplication by functions $f(\hat p\cdot e)$.  First note that any element $e \in H^3$ with $\hat p\cdot e \neq 0$, can be written as \beq\label{etz}e = \pm \Lambda_z^{-1}e_t\eeq where $\pm=\mathrm {sign}\,(\hat p\cdot e)$,
$\Lambda_z=\Lambda(A_{z,0})$ (cf.\ Eq.\ \eqref{E2}) and 
\beq e_t=(\sinh t,0,0,\cosh t).\eeq Explicitly,
\beq \exp(-t)=|e^0-e^3|=|\hat p\cdot e|,\quad z = -\frac{e^1-ie^2}{e^0-e^3} = -\frac{e^1-ie^2}{\hat p\cdot e} .\label{zlambda}\eeq
From Eq.\ \eqref{intrel} and \eqref{etz} we obtain \begin{equation}
u(\undertilde{e})=D^{(\frac j2,\frac k2)}(A_{z,0}^{-1})\, u( \undertilde{e_t}) 
\label{stintdef}
\end{equation}
so the intertwiner for $e$ is determined by those for the string directions $e_t$. We have therefore only to check that these latter intertwiners coincide with \eqref{intertwinere} up to multiplication by functions of $\hat p\cdot e_t$. To show this one notes that
${e_t}$ is invariant under rotations around the $x^3$-axis so by \eqref{intrel} $u(\undertilde{e_t})$ must fulfill
\begin{equation}
D^{(\frac j2,\frac k2)}(A_{0,\theta}^{-1})\, u(\undertilde e_t) = e^{ih\theta} u(\undertilde{e_t}).
\label{stintrel}
\end{equation}
Thus $u(\undertilde{e_t})$ has to be an eigenvector of $D^{(\frac{j}{2},\frac k{2})}(A_{0,\theta}^{-1})$ to the eigenvalue $e^{ih\theta}$. But a basis of such eigenvectors is given by
\begin{equation}\label{specialintertwiner}
\left[u_-^{\otimes m} \otimes_s u_+^{\otimes(j-m)}\right]\otimes \left[{u_-}^{\otimes\bar{m}}\otimes_s\overline{u_+}^{\otimes(k-\bar{m})}\right],
\end{equation}
where again $u_- = \binom{1}{0}$, $u_+ = \binom{0}{1}$ and $h = (\frac{j}{2}-m)-(\frac k{2}-\bar{m})$. 
Since 
\beq \undertilde {e_t}u_+=\mp \exp(-t)u_+=\mp (\hat p\cdot e_t)u_+\eeq
we see that $u(\undertilde{e_t})$ given by \eqref{intertwinere} is, indeed, unique up to a factor $f_i(\hat p\cdot e_t)$ for each basis element. By \eqref{intrel} the uniqueness carries over to all $e$ with $\hat p\cdot e\neq 0$. Finally, the analyticity properties of the intertwiners imply that they are already determined by their values in the neighborhood of any complex $e$ in the tuboid, so uniqueness holds also for $\hat p\cdot e=0$.

\subsubsection{A simple Example}
As an example consider the intertwiner for the 4-vector representation $D^{(\frac{1}{2},\frac{1}{2})}$ and helicity $h=0$. The equation $h=(\frac{j}{2}-m)-(\frac k{2}-\bar{m})$ implies $m = \bar{m}$.
Together with the inequalities $m\leq j = 1$ and $\bar{m}\leq k = 1$ one obtains two possibilities:
\begin{enumerate}[a)]
\item $m=1$, $\bar{m}=1$
\item $m=0$, $\bar{m}=0$
\end{enumerate}
The corresponding intertwiners are:
\begin{enumerate}[a)]
\item $u_-\otimes {u_-} = \begin{pmatrix}1&0\\0&0\end{pmatrix} = \undertilde{\hat p}$ \\
This leads to a vector-intertwiner of the form $u^\mu \propto \hat p^\mu$. This is just the point-like intertwiner corresponding to a vector field $a^\mu = \partial^\mu \Phi$, where $\Phi$ is the scalar field for helicity $0$.
\item $u_+(\undertilde{e})\otimes \overline{u_+(\undertilde{e})} \propto \undertilde{\hat p} + (e^0-e^3)\undertilde{e} = \undertilde{\hat p} + (\hat p\cdot e)\undertilde{e}$ \vspace{0.5em} \\
$\Rightarrow u^\mu \propto {\hat p}^\mu + (\hat p\cdot e)e^\mu$
\end{enumerate}
From these two possibilities it follows that a general vector-intertwiner for $h=0$ and $p=\hat p$ has the form
\begin{equation}
u^\mu(e) = f(\hat p\cdot e) {\hat p}^\mu + g(\hat p\cdot e) e^\mu
\end{equation}
and for general $p$ this means according to \eqref{upe}
\beq
u^\mu(p,e) = f( p\cdot e) {p}^\mu + g(p\cdot e) e^\mu.
\eeq
\subsection{Minimal intertwiners}
For a given $h$ we call the $D^{(\frac j2,\frac k2)}$-intertwiners with $(j+k)/2=|h|$ the {\it minimal intertwiners} (for this particular $h$). The name is justified because in this case there is only one independent intertwiner (up to multiplication by $f(p\cdot e)$) for each $(j,k)$. Clearly there are $2|h|+1$ minimal intertwiners for each $h$, corresponding to $j=2|h|-k$, $k=0,\dots,2|h|$. For $h\geq 0$ these are
\beq\label{minimal1} u(\undertilde e,h)=u_+(\undertilde e)^{\otimes j}\otimes u_-^{\otimes k}\eeq
and for $h<0$
\beq\label{minimal2} u(\undertilde e,-|h|)=u_-^{\otimes j}\otimes \overline{u_+(\undertilde e)}^{\otimes k}.\eeq
Independence of $\undertilde e$ holds if and only if either $j=0$, corresponding to $D^{(0,|h|)}$ and helicity $|h|$, or $k=0$, corresponding to $D^{(|h|,0)}$ and helicity $-|h|$. The passage from one minimal intertwiner to another  amounts to replacing a factor $u_+(\undertilde e)$ in \eqref{minimal1} or \eqref{minimal2} on one side of the tensor product by $u_-$ and vice versa.

For general $p$ on the mass shell the corresponding formulas are, according to \eqref{upe},
\beq\label{minimal1p} u(p, e,h)=\left[\undertilde e\,{B_p^*}^{-1}u_+\right]^{\otimes j}\otimes \left[ \bar B_p\,u_-\right]^{\otimes k}\eeq
and
\beq\label{minimal2p} u(p, e,-|h|)=\left[B_p\,u_-\right]^{\otimes j}\otimes\left[\undertilde{\bar  e}\,{B_p^T}^{-1}u_+\right]^{\otimes k}.\eeq

 Using the antisymmetric tensor 
$$\epsilon = \begin{pmatrix}0&1\\-1&0\end{pmatrix}$$ we obtain the relations
\begin{equation}\label{ref1}
(\undertilde{\hat p}\,\epsilon) u_+(\undertilde e)=(\undertilde{\hat p}\,\epsilon)\overline{u_+(\undertilde e)}=2 (\hat p\cdot e)u_-\end{equation}
and 
\beq\label{ref2}
{\undertilde{e}}\, \epsilon\, u_- = -{u_+(\undertilde{e})}, \quad \bar{\undertilde{e}}\, \epsilon\, u_- = -\overline{u_+(\undertilde{e})}.
\eeq
For general $p$ we denote
\beq u_+(p,e)=\undertilde e\,{B_p^*}^{-1}u_+, \quad u_-({p})=B_pu_-
\eeq
and, using that $A\,\epsilon\,A^T=\epsilon$ for all $A\in SL(2,\mathbb C)$, we obtain from \eqref{ref1} and \eqref{ref2} 
\beq\label{ref11} (\undertilde p\,\epsilon)\overline{u_+(p,e)}=(p\cdot e)u_-({p}),\qquad \undertilde e\,\epsilon\,u_-({p})=u_+(p,e).
\eeq
Thus,  one can move from one minimal intertwiner to another by applying either $(p\cdot e+i0)^{-1}(\undertilde{p}\,\epsilon)$,  or ${\undertilde{e}}\, \epsilon$ (resp.\ $\bar{\undertilde{e}}\, \epsilon$) to an appropriate factor in \eqref{minimal1p} or \eqref{minimal2p}. Here $(p\cdot e+i0)^{-1}$ denotes the distributional limit from the forward tuboid as the imaginary part of $e$ tends to zero. 

We note that in $x$-space a factor $\undertilde p\,\epsilon$ corresponds to the differential operator $-i\undertilde\partial\,\epsilon$, while $(p\cdot e+i0)^{-1}$ amounts to an integration along an infinite ray in the direction of $e$ starting at $x$.

For the Bose case, $h\in\mathbb Z$,  one can in particular choose intertwiners for the symmetric representation $D^{(\frac{|h|}{2},\frac{|h|}{2})}$, which can be used to construct self-adjoint fields according to Section 2.4. Indeed, since $u_-$ is independent of $e$ and $u_+(-\undertilde e) = -u_+(\undertilde e)$ one can always arrive at intertwiners satisfying the relation $\overline{u(p,e,h)} = u(p,-e,-h)$ required for a self-adjoint field by chosing the function $f(p\cdot e)$ such that $f(p\cdot (-e)) = (-1)^h f(p\cdot e)$. Other minimal intertwiners for the given helicity can then be obtained by the elementary operations described above from this particular one. We shall return to this point in Section 4 below.

\subsection{$D^{(\frac{j+b}{2},\frac{b}{2})}$- from $D^{(\frac{j}{2},0)}$-intertwiners}
Consider again the intertwiner $u^{(0)}(\undertilde{e}) := f(\hat p\cdot e) \undertilde{\hat p} + g(\hat p\cdot e) \undertilde{e}$ for helicity 0 as discussed in 3.2.1. above. It satisfies the equation
\begin{equation*}
D^{(\frac{1}{2},\frac{1}{2})}(A^{-1}) u^{(0)}(A\undertilde{e}A^\dagger) = u^0(\undertilde{e}).
\end{equation*}
For a general intertwiner $u(\undertilde{e})$ for a representation $D^{(\frac{j}{2},\frac k{2})}$ and helicity $h$ one can now take the tensor product with $u^{(0)}(\undertilde{e})$,
\begin{equation*}
\hat{u}(\undertilde{e}) := u(\undertilde{e})\otimes u^{(0)}(\undertilde{e}),
\end{equation*}
and symmetrize with respect to the dotted and undotted indices respectively. This $\hat{u}(\undertilde{e})$ is an intertwiner for the same helicity $h$, but for the representation $D^{(\frac{j+1}{2},\frac{k+1}{2})}$. 
The field defined by the new intertwiner still creates particles with helicity $h$ from the vacuum, but it now transforms according to the $D^{(\frac{j+1}{2},\frac{k+1}{2})}$ representation.

This can be generalized as follows. For a general intertwiner $u(\undertilde{e})$ for $D^{(\frac{j}{2},\frac k{2})}$ and helicity $h$ take an intertwiner $u^{(0)}(\undertilde{e})$ for representation $D^{(\frac{b}{2},\frac{b}{2})}$ and helicity $h=0$ from the $b+1$ dimensional intertwiner space (Note that $\min\{b,b\} + 1 = b + 1 - |h| = b+1$). By taking again the tensor product $u(\undertilde{e})\otimes u^{(0)}(\undertilde{e})$ and symmetrizing w.r.t the dotted and undotted indices one gets a new intertwiner $\hat{u}(\undertilde{e})$ for the same helicity, but for the representation $D^{(\frac{j+b}{2},\frac{k+b}{2})}$. \\
In this way one can create a field for $D^{(\frac{j+b}{2},\frac{b}{2})}$ from a field for $D^{(\frac{j}{2},0)}$ by tensoring it $b$ times with $\undertilde{\partial}$ or $\undertilde{e}$ (or both) and symmetrizing afterwards.

\subsection{The general structure of the intertwiners}

The conclusion that can be drawn from the last two subsections is that one can construct an intertwiner for every helicity $h$, transforming according to an arbitrary representation $D^{(\frac{j}{2},\frac k{2})}$ (as long as $|h| \leq  \frac{j+k}{2}$ is fulfilled), by starting with one of the minimal intertwiners discussed in 3.3 above and consecutively applying these elementary operations:
\begin{itemize}
\item Contraction of a dotted index with $\undertilde{p}_{\alpha\dot{\sigma}}\epsilon^{\dot{\sigma}\dot{\beta}}$
\item Contraction of an undotted index with $\bar{\undertilde{e}}_{\dot{\beta}\alpha}\epsilon^{\alpha \rho}$
\item Tensoring with an intertwiner $u^{(0)}({p,e})$ for helicity $h=0$ and subsequent symmetrization.
\end{itemize}
The intertwiners $u^{(0)}$ for helicity 0 are not unique, but by choosing a basis $u_i^{(0)}$  and taking the linear combination $\sum_{i=1}^d f_i(p\cdot e) u_i^0$ one can arrive in this way at all possible intertwiners for a given helicity and representation. 

Let's sketch this construction starting with the pointlike field for helicity $h=-|h|$, transforming according to $D^{(|h|,0)}$, with the corresponding intertwiner $(u_-)^{\otimes 2|h|}$ for $p=\hat p$. To get from this an intertwiner for the same helicity and representation $D^{(\frac{j}{2},\frac k{2})}$ one has to take in the first step a basis $u_i^{(0)}$ for the intertwiner space to $D^{(\frac{b}{2},\frac{b}{2})}$ and $h=0$. By taking linear combinations and tensoring them with $(u_-)^{\otimes 2|h|}$ one gets an intertwiner for $D^{(|h|+\frac{b}{2},\frac{b}{2})}$,
\begin{equation*}
\sum_{i=1}^{b+1} f_i(\hat p \cdot e)\, u_i^{(0)}\otimes (u_-)^{\otimes 2|h|} =: \hat{u},
\end{equation*}
where $b=\frac{j+k}{2}-|h|$. 

In the next step one uses the contractions described above to construct from this an intertwiner for $D^{(|h|+\frac{b+a}{2},\frac{b-a}{2})}$, where $a = \frac{j-k}{2}-|h|$ which can of course also be negative. To take this into account define $A := \Theta(a) |a|$ and $B := \Theta(-a) |a|$, where $\Theta$ is the Heaviside step function. The final intertwiner for $D^{(\frac{j}{2},\frac k{2})}$ and $h=-|h|$ then becomes
\begin{equation*}
(\undertilde{\hat p}_{\rho_1\dot{\sigma}_1}\epsilon^{\dot{\sigma}_1\dot{\beta}_1}... \undertilde{\hat p}_{\rho_A\dot{\sigma}_A}\epsilon^{\dot{\sigma}_A\dot{\beta}_A}) (\bar{\undertilde{e}}_{\dot{\gamma}_1 \delta_1}\epsilon^{\delta_1 \alpha_1}... \bar{\undertilde{e}}_{\dot{\gamma}_B \delta_B}\epsilon^{\delta_B \alpha_B})\, \hat{u}_{\alpha_1...\alpha_{2|h|+b}\dot{\beta}_1...\dot{\beta}_{b}} .
\end{equation*}
Of course only one of $A$ and $B$ can be non-zero, so one only needs to contract with either $\undertilde{\hat p}\epsilon$ or $\bar{\undertilde{e}}\epsilon$. This is in fact the most general intertwiner, because the degeneracy for $D^{(\frac{j}{2},\frac k{2})}$ and $|h|$ is $d=|h|+\frac{b+a}{2}+\frac{b-a}{2}+1-|h| = b+1$ which is the same as for the intertwiners $u_i^0$ above.

\section{String Localized Potentials for Point Localized fields}
\subsection{The case $h\in\mathds{Z}$}
We first discuss the bosonic case, $h \in \mathds{Z}$. As discussed in Section 3.3 one of the minimal intertwiners for 
helicity $h=-|h|$ and $p=\hat p$ is
\begin{equation}\label{orgint}
u_{\alpha_1...\alpha_{2|h|}}^{(F-)} = (u_-)_{\alpha_1}...(u_-)_{\alpha_{2|h|}},
\end{equation}
which is independent of $e$. This intertwiner transforms w.r.t.\ the representation $D^{(|h|,0)}$. A corresponding intertwiner for helicity $+|h|$, transforming w.r.t.\  $D^{(0,|h|)}$ is 
\begin{equation}\label{orgint_conj}
u_{\dot{\beta}_1...\dot{\beta}_{2|h|}}^{(F+)}= (u_-)_{\dot{\beta}_1}...(u_-)_{\dot{\beta}_{2|h|}}.
\end{equation}
Via the formulas \eqref{upe}, \eqref{vfromu} and \eqref{fieldansatz} these intertwiners define free, point localized Wightman fields, $F^{-}_{\alpha_1...\alpha_{2|h|}}(x)$ and 
$F^{+}_{\dot{\beta}_1...\dot{\beta}_{2|h|}}(x)$.  A self-adjoint field can be obtained by forming their sum that transforms according to $D^{(|h|,0)}\oplus D^{(0,|h|)}$, cf. Section 2.4.

Consider now the intertwiners 
\begin{equation}\label{orgint2}
u^{(A-)}_{\alpha_1...\alpha_{|h|}\dot{\beta}_1...\dot{\beta}_{|h|}}(\undertilde e) = \frac{1}{(\hat p\cdot e+i0)^{|h|}}\,(u_-)_{\alpha_1}...(u_-)_{\alpha_{|h|}} (\overline{u_+(\undertilde e)})_{\dot{\beta}_1}...(\overline{u_+(\undertilde e)})_{\dot{\beta}_{|h|}},
\end{equation}
and 
\begin{equation}\label{orgint2_conj}
u^{(A+)}_{\alpha_1...\alpha_{|h|}\dot{\beta}_1...\dot{\beta}_{|h|}}(\undertilde e) = \frac{1}{(\hat p\cdot e+i0)^{|h|}}\,(u_+(\undertilde e))_{\alpha_1}...(u_+(\undertilde e))_{\alpha_{|h|}} (u_-)_{\dot{\beta}_1}...(u_-)_{\dot{\beta}_{|h|}},
\end{equation}
The former is an intertwiner for $D^{(\frac{|h|}{2},\frac{|h|}{2})}$ and $h=-|h|$, the latter for $D^{(\frac{|h|}{2},\frac{|h|}{2})}$ and $h=+|h|$.

From \eqref{ref1} it now follows that
\begin{equation}\label{newint}
{u}^{(F-)}_{\alpha_1...\alpha_{|h|} \rho_1...\rho_{|h|}}= (\undertilde{\hat p}_{\rho_1\dot{\sigma}_1}\epsilon^{\dot{\sigma}_1\dot{\beta}_1} ... \undertilde{\hat p}_{\rho_{|h|}\dot{\sigma}_{|h|}}\epsilon^{\dot{\sigma}_{|h|}\dot{\beta}_{|h|}}) u^{(A-)}_{\alpha_1...\alpha_{|h|} \dot{\beta}_1...\dot{\beta}_{|h|}}(\undertilde e)
\end{equation}
and
\begin{equation}\label{newint2}
{u}^{(F+)}_{\dot{\sigma}_1...\dot{\sigma}_{|h|} \dot{\beta}_1...\dot{\beta}_{|h|}}= (\undertilde{\hat p}_{\dot{\sigma}_1{\rho}_1}\epsilon^{{\rho}_1\alpha_1} ... \undertilde{\hat p}_{\dot{\sigma}_{|h|}{\rho}_{|h|}}\epsilon^{{\rho}_{|h|}\alpha_{|h|}}) u^{(A+)}_{\alpha_1...\alpha_{|h|} \dot{\beta}_1...\dot{\beta}_{|h|}}(\undertilde e)
\end{equation}
The intertwiners $u^{(F,\pm)}({p})$ and $u^{(A,\pm)}({p},e)$ for arbitrary $p$ on the mass shell are defined in accord with \eqref{minimal1p} and \eqref{minimal2p} and the replacement of the pre-factor $(\hat p\cdot e+i0)^{-|h|}$ by
$(p\cdot e+i0)^{-|h|}$. From Eqs. \eqref{newint}, \eqref{newint2} and the corresponding equations with $p$ instead of $\hat p$, cf. \eqref{ref11}, we then see that the string localized fields $A^{\pm}(x,e)$ defined by the intertwiners $u^{(A,\pm)}({p},e)$ and Eq. \eqref{vfromu} are, indeed, {\it potentials} for the fields $F^{\pm}(x)$ because in $x$-space these equations amount to
\begin{equation}\label{potential}
(-i)^{|h|}(\undertilde{\partial}_{\rho_1\dot{\sigma}_1}\epsilon^{\dot{\sigma}_1\dot{\beta}_1} ... \undertilde{\partial}_{\rho_{|h|}\dot{\sigma}_{|h|}}\epsilon^{\dot{\sigma}_{|h|}\dot{\beta}_{|h|}})\, A^{\pm}(x,e)_{\alpha_1...\alpha_{|h|} \dot{\beta}_1...\dot{\beta}_{|h|}} = F^{\pm}(x)_{\alpha_1...\alpha_{|h|} \rho_1...\rho_{|h|}}.
\end{equation}
Moreover, 
\beq \overline{u^{(A+)}(p,e)}=u^{(A-)}(p,-e)\eeq
 so the field $A=A^++A^-$ is self adjoint (cf. Section 2.4) and it is a potential for $F=F^++F^-$.

A further point to note is that the behavior of the two point function of the potential $A$ for large $p$ is independent of the helicity because although \eqref{minimal1p} and \eqref{minimal2p} grow as $|p|^{|h|}$ for large $p$ if $j=k=|h|$, this behavior is compensated by the factor $(p\cdot e+i0)^{-|h|}$ in $u^{A\pm}(p,e)$.

Equations \eqref{newint} and \eqref{newint2} can be inverted by making use of Eq.\ \eqref{ref2} to obtain
\begin{equation}\label{inversion1}
(\undertilde{\bar e}_{\dot{\beta}_1{\sigma}_1}\epsilon^{{\sigma}_1\rho_1} ... \undertilde{\bar e}_{\dot{\beta}_{|h|}{\sigma}_{|h|}}\epsilon^{{\sigma}_{|h|}\rho_{|h|}}) 
u^{(F-)}_{\alpha_1...\alpha_{|h|}
\rho_1...\rho_{|h|}
}=(\hat p\cdot e+i0)^{|h|}{u}^{(A-)}_{\alpha_1...\alpha_{|h|} \dot{\beta}_1...\dot{\beta}_{|h|}}(\undertilde e)
\end{equation}
and
\begin{equation}\label{inversion2}
\hskip1.6cm(\undertilde{e}_{\alpha_1\dot{\sigma}_1}\epsilon^{\dot{\sigma}_1\dot{\rho}_1} ... \undertilde{e}_{\alpha_{|h|}\dot{\sigma}_{|h|}}\epsilon^{\dot{\sigma}_{|h|}\dot{\rho}_{|h|}}) u^{(F+)}_{\dot\rho_1\dots\dot\rho_{|h|} \dot{\beta}_1...\dot{\beta}_{|h|}}=(\hat p\cdot e+i0)^{|h|}{u}^{(A+)}_{\alpha_1\dots\alpha_{|h|} \dot{\beta}_1\dots\dot{\beta}_{|h|}}(\undertilde e).
\end{equation}
In $x$-space it means that the potentials can be written as integrals over the fields:
\beq
\int_0^\infty\cdots \int_0^\infty (\mathbf 1\otimes(\undertilde{\bar e}\,\epsilon)^{\otimes |h|})F^{-}(x+(t_1+\cdots+t_{|h|})e)dt_1\cdots dt_{|h|}=A^{-}(x,e)
\eeq
\beq
\int_0^\infty\cdots \int_0^\infty ((\undertilde{e}\,\epsilon)^{\otimes |h|}\otimes\mathbf 1)F^{+}(x+(t_1+\cdots+t_{|h|})e)dt_1\cdots dt_{|h|}=A^{+}(x,e).
\eeq

\subsection{The case $h\in\mathds{Z}+1/2$}
For the case of half-integer helicity $|h|\in\frac{\mathds{N}}{2}$, this construction has to be modified slightly, because then there is no representation ``$D^{(\frac{|h|}{2},\frac{|h|}{2})}$''. Instead take the intertwiner
\begin{equation}
\left( \frac{1}{(\hat p\cdot e+i0)^{|h|-\frac{1}{2}}} \right)\, \left( (u_-)^{\otimes(|h|+\frac{1}{2})}\otimes (\overline{u_+(\undertilde{e})})^{\otimes(|h|-\frac{1}{2})}\right) =: \hat{u}(\undertilde{e}) ,
\end{equation}
which is an intertwiner for $D^{(\frac{|h|}{2}+\frac{1}{4},\frac{|h|}{2}-\frac{1}{4}]}$ and $h=-|h|$. (This makes sense because $|h|\pm\frac{1}{2}\in\mathds{N}$ for $|h|$ half-integer.) \\
This intertwiner then is again a potential for $(u_-)^{\otimes 2|h|}$ in the sense that
\begin{equation*}
\left( \undertilde{\hat p}_{\rho_1\dot{\sigma}_1}\epsilon^{\dot{\sigma}_1\dot{\beta}_1} ... \right) \, \hat{u}(\undertilde{e})_{\alpha_1... \dot{\beta}_1 ...} = (u_-)_{\alpha_1}... (u_-)_{\rho_1} ...\, .
\end{equation*}
So the field strength $F(x)$ for helicity $h=-|h|$ and $D^{(|h|,0)}$ can be obtained by applying the derivative operator $\undertilde{\partial}_{\rho\dot{\sigma}}\epsilon^{\dot{\sigma}\dot{\beta}}$ a total of $(|h|-\frac{1}{2})$-times to the field $A(x,e)_{\alpha...\dot{\beta}...}$ for $D^{(\frac{|h|}{2}+\frac{1}{4},\frac{|h|}{2}-\frac{1}{4})}$.
The $h=+|h|$ case can be dealt with in the same way by just taking the complex conjugates of the intertwiners.


\subsection{Vector and tensor potentials}

For the Bose case, $h\in\mathbb Z$, the fields and the potentials can also be written in vector or tensor notation instead of the spinor notation employed above. We state here the basic formulas.

For the  representation $D^{(|h|,0)} \oplus D^{(0,|h|)}$ one gets the (real) field strength tensors $F_{\mu_1\nu_1...\mu_{|h|}\nu_{|h|}}$ with intertwiners 
\begin{equation}\label{tensorintetwiners1}
\hat{e}_\pm(p)_{[\mu_1}p_{\nu_1]} ... \hat{e}_\pm(p)_{[\mu_{|h|}}p_{\nu_{|h|}]} =: u(p)_{\mu_1\nu_1...\mu_{|h|}\nu_{|h|}}\ , \eeq
with the polarization vectors
\beq\label{polarization}
 \hat{e}_\pm(p) = \Lambda(B_p) (0,1,\pm i,0) 
\end{equation}
where $\pm$ indicates the sign of the helicity and the square bracket antisymmetrization w.r.t.\ the four-vector indices $\mu_i,\nu_i$. Note that the norm of  $\hat{e}_\pm({p})$ is bounded independently of $p$ as can be seen by inserting \eqref{boost} in \eqref{polarization}.

For example for ${|h|}=1$ one gets the electromagnetic field strength $F_{\mu\nu}(x)$. A vector potential $A_\mu(x)$ for the field strength should satisfy 
\begin{equation*}
\partial_\mu A_\nu - \partial_\nu A_\mu = F_{\mu\nu}
\end{equation*}
and using the polarization vectors $\hat{e}_\pm$ as intertwiners for $A_\mu$ this relation can be fulfilled. The resulting field, however, does not transform covariantly under the Poincaré group with the vector representation $D^{(\frac 12,\frac 12)}$ the reason being that this representation and ${|h|}=1$ violate $\frac{k-j}{2}=h$.

A possible solution to this problem, proposed in \cite{msy05},  is to use a string-localized vector potential $A_\mu(x,e)$, for which an appropriate intertwiner exists, according to the above considerations. It can be checked directly that the vector-intertwiner
\begin{equation}
u_\pm^\mu(e) := f(\hat p\cdot e)[(\hat p\cdot e) \hat{e}_\pm^\mu({\hat p}) - {\hat p}^\mu (\hat{e}_\pm({\hat p})\cdot e)]
\label{vectorint}
\end{equation}
leads in spinor notation to the same expression as $f(\hat p\cdot e) [u_\pm\otimes\overline{u_\mp(\undertilde{e})}]$ (up to an unimportant factor). In order to get a field $A_\mu$ that is a potential for $F_{\mu\nu}$, the intertwiner has to satisfy
\begin{equation*}
p_\mu u_{\pm \nu}(e) - p_\nu u_{\pm\mu}(e) \overset{!}{=} u_{\mu\nu}(p),
\end{equation*}
which has to be independent of $e$ in particular. To achieve this the function $f(p\cdot e)$ has to be chosen as 
$f(p\cdot e) = (p\cdot e + i0)^{-1}$.
The resulting vector field then satisfies $\partial_\mu A_\nu(x,e) - \partial_\nu A_\mu(x,e) = F_{\mu\nu}(x)$. 

For higher tensor fields $A(x,e)_{\mu_1...\mu_{|h|}}$ one can just take the tensor product of the intertwiners \eqref{vectorint}, with a function $f(p\cdot e) = (p\cdot e + i0)^{-|h|}$. 
\begin{equation*}
u_\pm(p,e)_{\mu_1...\mu_{|h|}} = \frac{1}{(p\cdot e + i0)^{|h|}} \left[ \hat{e}_\pm(p)_{[\mu_1}p_{\nu_1]}...\hat{e}_\pm(p)_{[\mu_{|h|}}p_{\nu_{|h|}]}  \right] e^{\nu_1}...e^{\nu_{|h|}}
\end{equation*}

This field is a potential for the field strength $F(x)$ in the sense that the expression
\begin{equation*}
\partial_{\mu_1}...\partial_{\mu_{|h|}} A(x,e)_{\nu_1...\nu_{|h|}},
\end{equation*}
when antisymmetrized in every index pair $\mu_k$, $\nu_k$, is equal to the field strength tensor $F(x)_{\mu_1...\mu_{|h|}\nu_1...\nu_{|h|}}$ (and especially independent of $e$). \\
The tensor potential $A(x,e)$ has further convenient properties:
\begin{itemize}
\item Total symmetry
\item Generalized Lorentz condition: $\partial^{\mu_1} A(x,e)_{\mu_1...\mu_{|h|}} = 0$
\item Axial ``gauge'' condition: $e^{\mu_1} A(x,e)_{\mu_1...\mu_{|h|}} = 0$
\item Trace free: $\eta^{\mu_1 \mu_2}A(x,e)_{\mu_1 \mu_2...\mu_{|h|}} = 0$ \text{(with $\eta$ the Minkowski metric)} 
\end{itemize}

Another interesting example, besides the photon field, is the string-localized field for helicity $h = \pm 2$, describing hypothetical gravitons \cite{mund09}. The field $h(x,e)_{\mu\nu}$ describes the perturbation of the metric $g_{\mu\nu} = \eta_{\mu\nu} + h_{\mu\nu}$ and is a potential for the linearized (point-localized) Riemann tensor $R(x)_{\mu\nu\rho\sigma}$. This means that the classical relation between $R_{\mu\nu\rho\sigma}$ and the field $h_{\mu\nu}$ holds, i.e.
\begin{equation}
R(x)_{\mu\nu\rho\sigma} = \frac{1}{2}\left[\partial_\mu\partial_\rho h(x,e)_{\nu\sigma} + \partial_\nu\partial_\sigma h(x,e)_{\mu\rho} - \partial_\nu\partial_\rho h(x,e)_{\mu\sigma} - \partial_\mu\partial_\sigma h(x,e)_{\nu\rho}\right] ,
\end{equation}
which is a special case of the general relation between potentials and field strengths. The field $h(x,e)$ has all the desired properties, one demands a quantum field describing (linearized) gravity to have. Namely it is \emph{symmetric}, it satisfies the \emph{axial gauge condition} $e^\mu h(x,e)_{\mu\nu} = 0$ and the remaining properties $\partial^\mu h(x,e)_{\mu\nu} = 0$ and $h(x,e)^\mu_{\phantom{\mu}\mu} = 0$ are usually called \emph{harmonic gauge}. 


\section{The Case of Infinite Helicity}

For mass zero there exist besides the finite helicity representations also irreducible representations of the inhomogeneous $SL(2,\mathbb C)$ where the helicity is unbounded \cite{wigner}. In fact there are two classes of such representations. In the first class the helicities take all integer values, in the second class all half-integer values. The representations of the second class correspond to two-valued representations of the orthochronous Poincar\'e group. In both cases there is a continuum of inequivalent representations labelled by a parameter $\kappa^2>0$ (the value of the Pauli-Lubanski Casimir operator). 

String localized Bose fields corresponding to the first class of representations, where the helicity takes all integer values were constructed in \cite{msy04, msy05} and we recall this construction. Let $\mathcal H_\kappa$ be the Hilbert space of functions of $k\in\mathbb R^2$,
square integrable w.r.t. the measure $d\nu_\kappa(k)=\delta(|k|^2-\kappa^2)d^2k$. The representation of the little group $\widetilde {E(2)}$ on $\mathcal H_\kappa$ is defined by
\beq \label{infspinrep}(D_\kappa(A_{z, \theta})\varphi)(k)=e^{iz\cdot k}\varphi(R_\theta^{-1}k)
\eeq 
with $A_{z, \theta}$ as in \eqref{E2} ($z$ is here regarded as an element of $\mathbb R^2$) and $R_\theta$ a rotation by $\theta$. The one-particle Hilbert space is the space of $\mathcal H_\kappa$-valued functions of $p\in\mathbb R^4$, square integrable w.r.t. the measure $d\mu({p})=\Theta(p^0)\delta(p\cdot p)d^4p$, and the unitary transformation law on the one particle space is
\beq (U(a,A)\psi)({p})=e^{ip\cdot a}D_\kappa(R(p,A))\psi(\Lambda(A)^{-1}p)
\eeq
with $R(p,A)$ the Wigner rotation. The intertwiners are given by
\beq u^{(\gamma)}(p,e)(k)=e^{-i\gamma/2}\int d^2 z e^{iz\cdot k}(B_p\xi({z})\cdot e)^{\gamma}\eeq with ${\rm Re}\, \gamma <0$ and 
\beq \xi({z})=\left(\frac12 (|z|^2+1),z_1,-z_2,\frac12(|z|^2-1)\right).\eeq The corresponding string field is then defined in terms of the creation and annihilation operators $a^*(p,k)$ and $a(p,k)$ as
\beq\phi(x,e)=\int d\mu({p})\int d\nu_\kappa(k) [e^{ipx}u(p,e)(k)a^*(p,k)+ {\rm h.c.}].
\eeq

 The representation for unbounded half-integer helicity can be obtained by tensoring the representation \eqref{infspinrep} with the one-dimensional representation $A_{z, \theta}\mapsto e^{i\theta/2}$ for helicity $1/2$. The resulting representation of the little group is given on the same Hilbert space of $L^2$-functions on the circle with radius $\kappa$ as \eqref{infspinrep} but with the modified formula 
\beq (D_{\kappa,-}(A_{z, \theta})\varphi)(k)=e^{i\theta/2}e^{iz\cdot k}\varphi(R_\theta^{-1}k).\eeq
To obtain covariant fields one needs intertwiners $u^{(\gamma)}(p,e)_\alpha(k)$ that carry an additional spinor index $\alpha$. Such intertwiners are easily obtained as a product of an intertwiner \eqref{intertwinere} for helicity 1/2, e.g. $u_-$ that transforms with $D^{(\frac 12,0)}$. The complete formula is thus
\beq u^{(\gamma)}(p,e)_\alpha(k)=u^{(\gamma)}(p,e)(k)((B_p^{-1})u_-)_\alpha\eeq


\noindent{\bf Acknowledgements.} This work is supported by the project P22929--N16 funded by the Austrian Science Fund (FWF).

\end{document}